\documentclass[intlimits,twoside,a4paper]{article}

\usepackage[active]{srcltx}

\usepackage{amsmath,amssymb}
\usepackage{graphicx}
\usepackage{wrapfig}
\usepackage{xcolor}
\usepackage[T2A]{fontenc}
\usepackage[cp1251]{inputenc}

\usepackage{cmpj2}

\issue{2013}{16}{4}{43801}
\doinumber{10.5488/CMP.16.43801}

\title[Structure of cylindrical electric double layers]
{Structure of cylindrical electric double layers: Comparison
of density functional and modified Poisson-Boltzmann theories
with Monte Carlo simulations\thanks{It is a pleasure to dedicate
this paper to Dr. Myroslav Holovko on the occasion of his 70th
Birthday.}}

\author[V. Dorvilien \textsl{et al.}]{V. Dorvilien\refaddr{label1}, C.N. Patra\refaddr{label2}, L.B. Bhuiyan\refaddr{label1}, C.W. Outhwaite\refaddr{label3}}

\addresses{
\addr{label1}Laboratory of Theoretical Physics, Department of
Physics, Box 70377, University of Puerto Rico, \\ San Juan, Puerto
Rico 00936-8377, USA
\addr{label2}Theoretical Chemistry Section, Chemistry Group, Bhabha Atomic Research Centre, Mumbai 400085, India
\addr{label3}Department of Applied Mathematics, University of Sheffield, Sheffield S3 7RH, UK
}

\date{Received June 27, 2013}
\authorcopyright{V. Dorvilien, C.N. Patra, L.B. Bhuiyan, C.W. Outhwaite, 2013}

\begin{document}

\maketitle

\begin{abstract}
The structure of cylindrical double layers is studied using a modified
Poisson Boltzmann theory and the density functional approach. In the model double
layer, the electrode is a cylindrical polyion that is infinitely long, impenetrable,
and uniformly charged. The polyion is immersed in a sea of equi-sized rigid ions embedded in a
dielectric continuum. An in-depth comparison of the theoretically predicted
zeta potentials, the mean  electrostatic potentials, and the electrode-ion
singlet density distributions is made with the corresponding Monte Carlo
simulation data. The theories are seen to be
consistent in their predictions that include variations in ionic diameters, electrolyte
concentrations, and electrode surface charge densities, and are also capable of
well reproducing  some new and existing Monte Carlo results.

\keywords electric double layer, restricted primitive model, density profiles

\vspace{-0.1in}
\pacs 82.45.Fk, 61.20.Qg, 82.45.Gj
\end{abstract}
\section {Introduction}

Description of the interactions and correlations of large polyions
with the small, more mobile ions in the surrounding ionic
cloud is of significance in situations ranging from fundamental life
processes such as the transport of ions, water, and various
molecules across cell membranes,
flocculation in colloidal systems \cite{mandel1}, industrial
polyelectrolytes \cite{hara1}, and the native structure of DNA and
various proteins \cite{ha1,mhsfh1,js1}. The structure and
thermodynamics of all these systems manifest the complex behaviour of
small ions within the atmosphere.  A detailed understanding of the static
structural features is therefore central to assessing the relative effect
of various control parameters in such phenomena.

    The electrode (polyion) along with the neighbouring inhomogeneous ion
layer constitute the \emph{electric double layer} with the shape of
the polyion determining the geometry of the double layer. Over the
past few decades the planar double layer (PDL) has become synonymous with the
electric double layer as it has been the one to have been extensively investigated
through theoretical approaches, numerical simulation methods, and experimental
techniques (see for example, references \cite{hb,cher} for recent reviews). This notwithstanding,
other double layer systems, viz., the cylindrical double layer (CDL), the spherical double layer (SDL),
and ellipsoidal double layer (ESDL) have been coming under increasing scrutiny in
recent years.

A number of experimental techniques including the small angle x-ray and neutron scattering
\cite{bov96,zebobvm}, the optical imaging, and electrophoretic mobility measurements have been used
to probe the properties of electric double layers \cite{isr1,lr1}. Recent advances in theoretical approaches
and simulation methodologies have aided in the explanation of some of these experimental observations \cite{levin}.
In most of the theoretical studies, the polyion is generally modelled as an
infinitely long, hard, uniformly charged cylinder  with the small
ions being treated as charged hard spheres moving in a dielectric continuum. This
is the so-called primitive model of the CDL. Of the
theoretical studies mention ought to be made of the counterion condensation (CC)
theory \cite{mann2}, the classical Poisson-Boltzmann (PB) description \cite{stigter1,sh1}, and the
formal statistical mechanical approaches, for example, the integral equation theories
\cite{hend1,yfsl1}, the modified Poisson-Boltzmann theory (MPB) \cite{outh,bo1,bo2}, and the density
functional theory (DFT) \cite{py1,py2}. Parallel Monte Carlo
(MC) simulations in various forms \cite{mar1,mar2,nar1,ar1,dbbo1,dbbo2,zebobvm,bv1,bv2,lbz,vh1}
have provided \emph{exact} quantitative data for many model systems with regard to their structure and thermodynamics.
However, many of these simulations have involved the cylindrical cell model \cite{dbbo1,dbbo2,zebobvm,bv1,bv2,lbz,vh1} where
the polyion is at a finite, non-zero concentration. By contrast, in the case of the
CDL~--- our focus in this paper, the polyion is at infinite dilution, and in this situation
the early MC simulations \cite{mar1,mar2,nar1,ar1} were somewhat limited in their scope.
Indeed, in an earlier paper \cite{pb1} two of us tried to compare the DFT and the MPB
structural results for the primitive model CDL but the lack of detailed simulation data
proved a hindrance and precluded a detailed comparison.
The picture has now changed, however, with the availability of a new generation
of MC simulation results due to Goel {et al.} \cite{Goel}. This will
permit a critical evaluation of these theories relative to the benchmark of the
simulation data.

    The DFT and MPB approaches to the electric double layer theory
have come to be recognized as being two of the more successful ones
for the planar \cite{bo3}, spherical \cite{bo4,goel2}, and cylindrical \cite{pb1}
symmetries. In the PDL, in particular, both the theories have proved
useful in describing the capacitance behaviour including capacitance
anomaly at low temperatures \cite{boh1}. The DFT techniques have evolved
over the years since the initial applications to the PDL in the early 1990's
\cite{ttdsw1,tswd1,ttdsw2}. In recent years, a number of new formulations of DFT, coupled with
different statistical mechanical approaches, have emerged as robust methods to study
the systems involving correlations like the electric double layer \cite{lw06}.
A partially perturbative DFT procedure \cite{py1,pg3} is adopted in the present work
where the hard sphere part is determined through the weighted density
approach (WDA) of Denton and Ashcroft \cite{da1}, while the
electrical contribution is a perturbation on
the corresponding bulk electrolyte.
The essential idea of the MPB theory is to incorporate within a potential
formulation, the important missing elements in the classical PB
formulation, namely, the inter-ionic correlations and the ionic exclusion
volume effects. For the CDL, an MPB equation was first obtained by
Outhwaite \cite{outh}. Numerical solutions were later developed
by Bhuiyan and Outhwaite \cite{bo1,bo2} with some limited comparison
with the hypernetted chain/mean spherical approximation \cite{lc1,glh1}.

    The first detailed comparative study of the DFT and the MPB structural results
relative to MC simulations was undertaken by Bhuiyan and Outhwaite \cite{bo3}
for the PDL. The two theories were seen to be remarkably consistent over a
wide range of physical parameters probed through the simulations \cite{bfhs1}.
A similar consistency between the theories was later observed by the same
authors \cite{bo4} with regard to structure in a SDL. An interesting result
of the latter work is the charge inversion phenomenon observed in the MC
simulations \cite{tn1} for higher valencies, which was also reproduced by
both the theories to a very good accuracy. It is thus natural to wonder if
such trends will carry over to cylindrical symmetries.

    In this paper we will explore the DFT and MPB theories for a
primitive model CDL with a particular emphasis on the comparative behaviour
of zeta potentials, density and mean electrostatic profiles
vis-a-vis the Monte Carlo data. The zeta potentials are useful indicators
of the capacitance characteristics of double layers and such comparison
of these two theories in this regard has not appeared in the literature.
Although our focus will be on DFT and MPB calculations at
different physical states, we also intend to do MC simulations for
new states.

\section{Model and methods}

As indicated in the previous section, in the model double layer system
treated here, the polyion is mimicked by an infinitely long, non-polarizable,
hard cylinder with a uniform surface charge density. The polyion is bathed
by a restricted primitive model (RPM) electrolyte (equi-sized charged hard spheres
moving in a dielectric continuum). The polyion surface charge density $\sigma $
is related to the axial charge $\xi$ through
\begin{equation}
\sigma = \frac{e}{2\pi Rb} = \frac{2\epsilon _{0}\epsilon _\textrm{r}}{\beta eR}\xi\,,
\end{equation}
where $e$ is the fundamental charge, $b$ is the monomer length
(length per unit charge), $\beta =(k_\textrm{B}T)^{-1}$,
$k_\textrm{B}$ being the Boltzmann's constant, and $T$ the temperature.
The continuum solvent is characterized by the relative permittivity
$\epsilon _\textrm{r}$ ($\epsilon _{0}$ is the vacuum permittivity), which is taken to be
$\epsilon _\textrm{r}= 78.358$, and
$R$ is the radius of the polyion taken as $R=  8 \times 10^{-10}$~m. Following
the previous work of Patra and Bhuiyan \cite{pb1} we have
set $T=298.15$~K,  $b=1.7\times 10^{-10}$~m, and in most
cases $\xi = 4.2$ ($\sigma = 0.187$~C/m$^{2}$),
which are the accepted values for a double-stranded
DNA \cite{ar1}. The other variable physical quantities will be described
in the Results section. Denoting the charge of a small ion of species ${i}$
by $q_{i} = Z_{i}e$, with $Z_{i}$ being the ionic valency, the
ion-ion interaction potential in the Hamiltonian can be written as
\begin{equation}
 u_{ij}(r)=\left\{
\begin{array}{ll}
 \infty, & \hbox{$r<a$}, \\[1ex]
\displaystyle
\frac{q_{i} q_{j}}{4\pi\epsilon _{0}\epsilon _\textrm{r}r}, & \hbox{$r>a$,}
\end{array}
\right.
\end{equation}
where $r$ is the distance between a pair of ions and $a$ is the common
diameter of the ions of the bathing electrolyte.
The bare interaction between an ion of species $i$
and the polyion is given by
\begin{equation}
         u_{i}(r_{i})=v_{i}(r_{i})+w_{i}(r_{i}),
        \end{equation}
where $v_{i}(r_{i})$ and $w_{i}(r_{i})$ are the non-electrostatic and electrostatic (Coulombic)
parts of the ion-polyion potential with $r_{i}$ being radial distance of ion $i$ from
the polyion axis.  The non-electrostatic
contribution  is a hard-core potential
 \begin{equation}
 v_{i}(r_{i})=\left\{
\begin{array}{ll}
 \infty, & \hbox{$r_{i}<R+a/2$}, \\
0, & \hbox{$r_{i}>R+a/2$}.
\end{array}
\right.
\end{equation}
The electrostatic part $w_{i}(r)$ is given by
\begin{equation}
w_{i}(r)
=\left\{
\begin{array}{ll}
  \displaystyle  -\frac{eq_{i}}{2\pi\epsilon _{0}\epsilon _\textrm{r} b} {\ln}(r_{i}),
     &   \hbox{$r_{i}\geqslant R+a/2$},   \\
     \infty, & \hbox{otherwise}.
\end{array}
      \right. \label{ups1}
\end{equation}

   The above double layer model was solved using the DFT and MPB techniques.
The development of these theories has been chronicled elsewhere in the literature
and will not be repeated here. For a summary of the principal equations
that are used in the present calculations we refer the reader again to the
earlier work by Patra and Bhuiyan \cite {pb1}.  The MC simulations were done in the canonical ensemble
using the standard Metropolis algorithm and have been described by Goel {et al.}
\cite{Goel}.

\section{Results and discussion}

\begin{figure}[!t]
\centerline{
\includegraphics[width=0.45\textwidth]{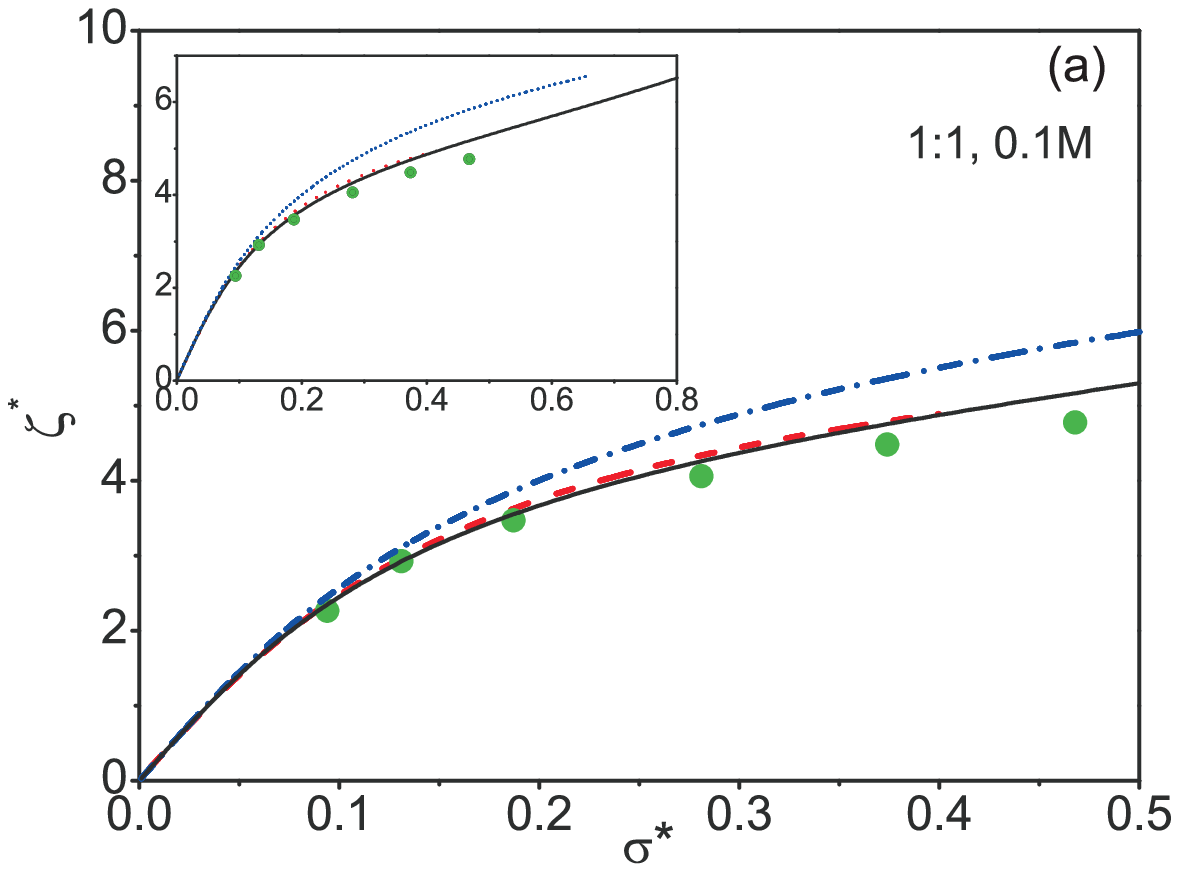}
\hspace{5mm}
\includegraphics[width=0.45\textwidth]{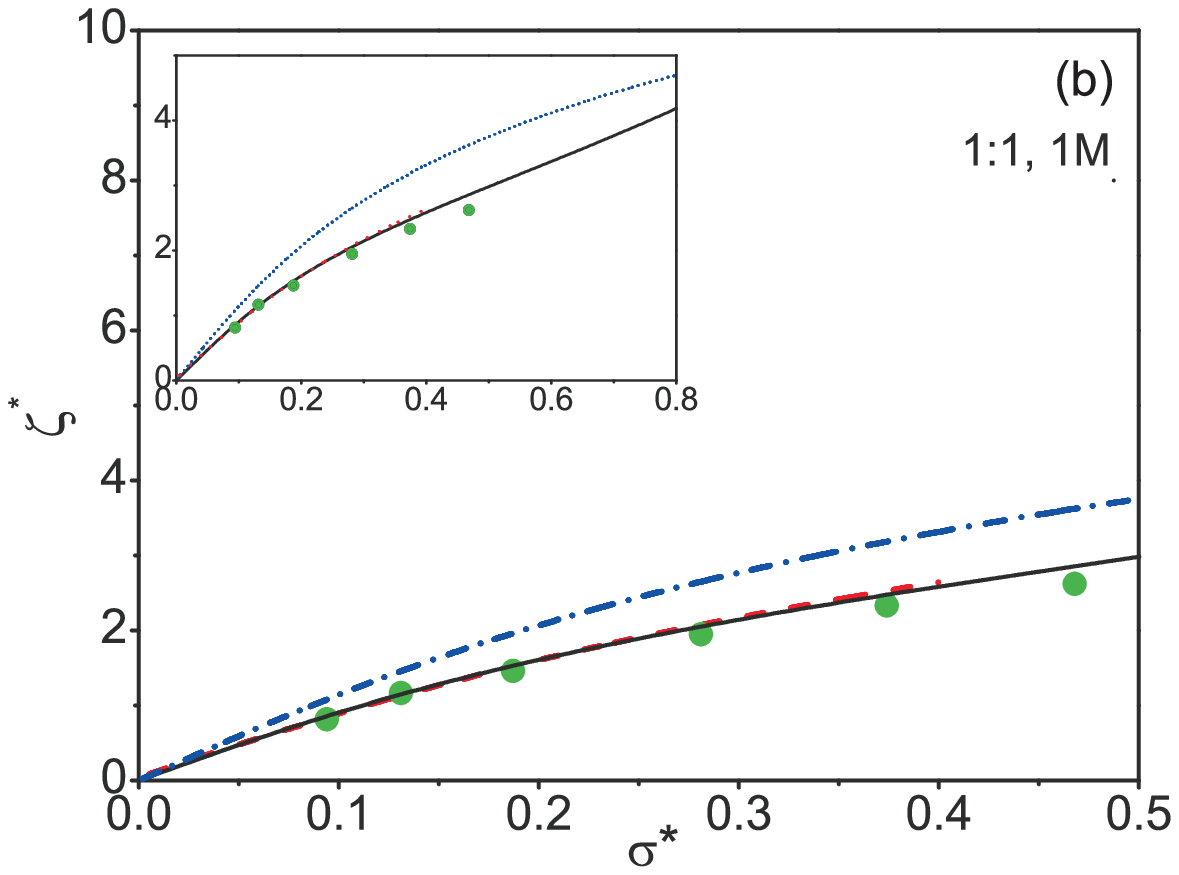}
}\vspace{1ex}
\centerline{
\includegraphics[width=0.45\textwidth]{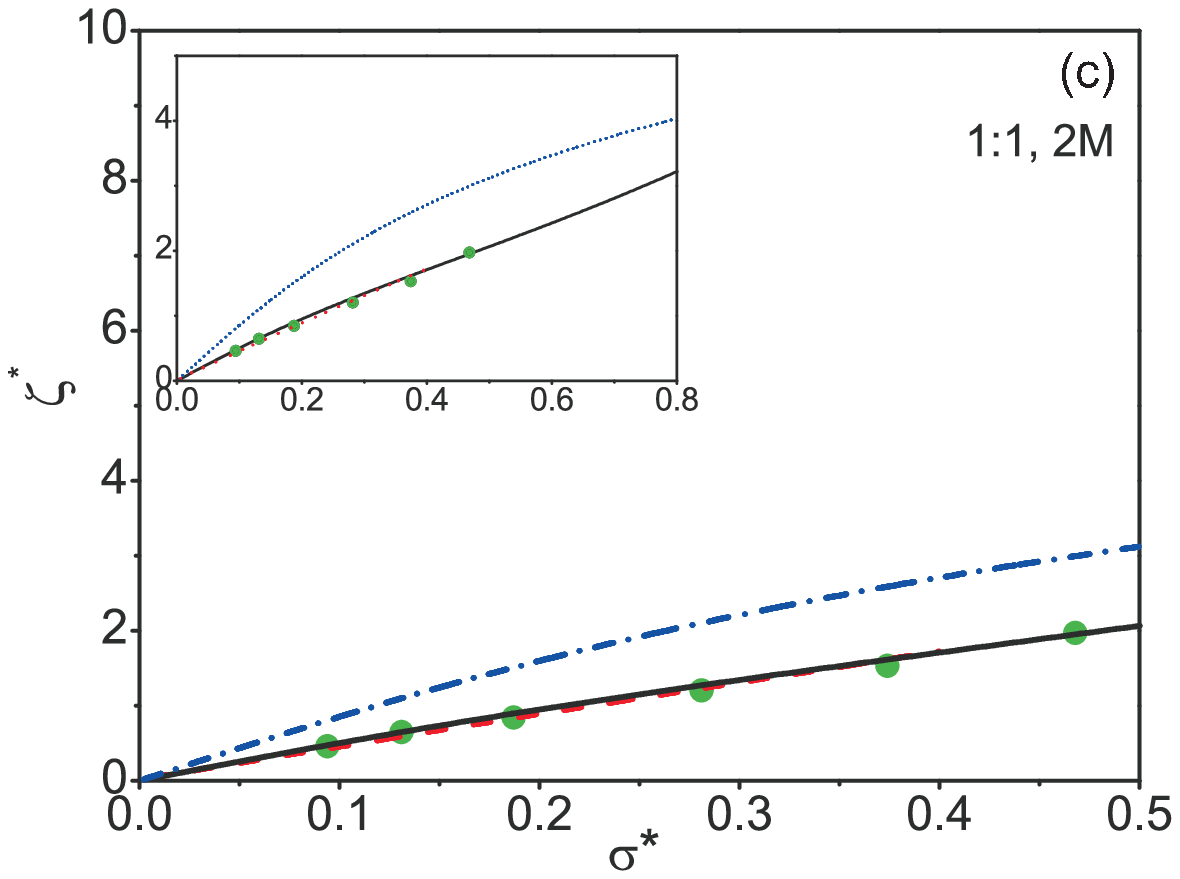}
}
\caption{(Color online) Reduced zeta potential $\zeta ^{*}$ [$=\psi ^{*}(R/a+1/2)$] versus
reduced surface charge density $\sigma ^{*}$ for a 1:1 electrolyte
at solution concentration (in mol/dm$^{3}$ units) 0.1~M (panel a),
1~M (panel b), and 2~M (panel c) in a RPM cylindrical double layer.
The symbols represent MC data, while the solid
line represents the MPB results, the dashed line represents the DFT results,
and the dash-dotted line shows the PB results. The polyion radius is $R = 8 \times 10^{-10}$~m,
and the ionic diameter $a = 4 \times 10^{-10}$~m. MC data from reference \cite{Goel}.}
\label{f1}
\end{figure}

\begin{figure}[!t]
\centerline{
\includegraphics[width=0.45\textwidth]{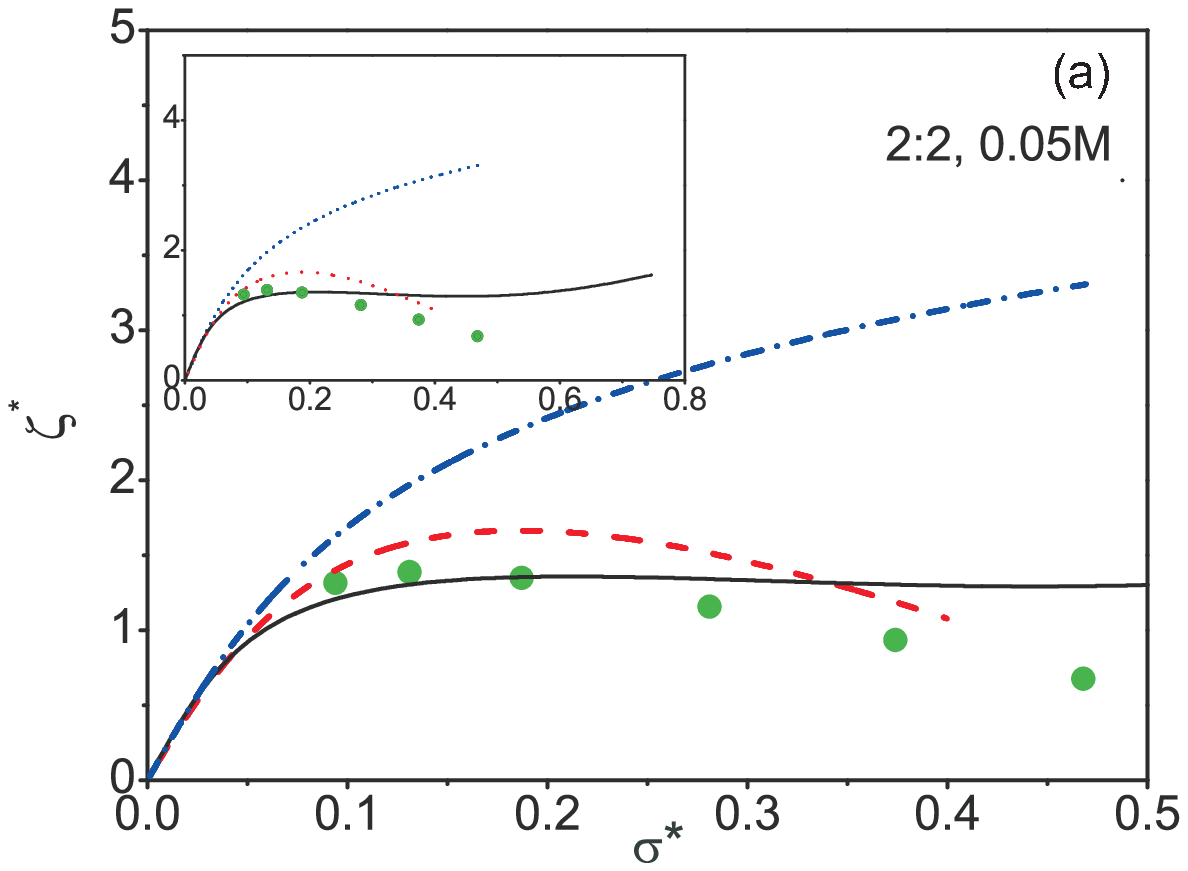}
\hspace{5mm}
\includegraphics[width=0.45\textwidth]{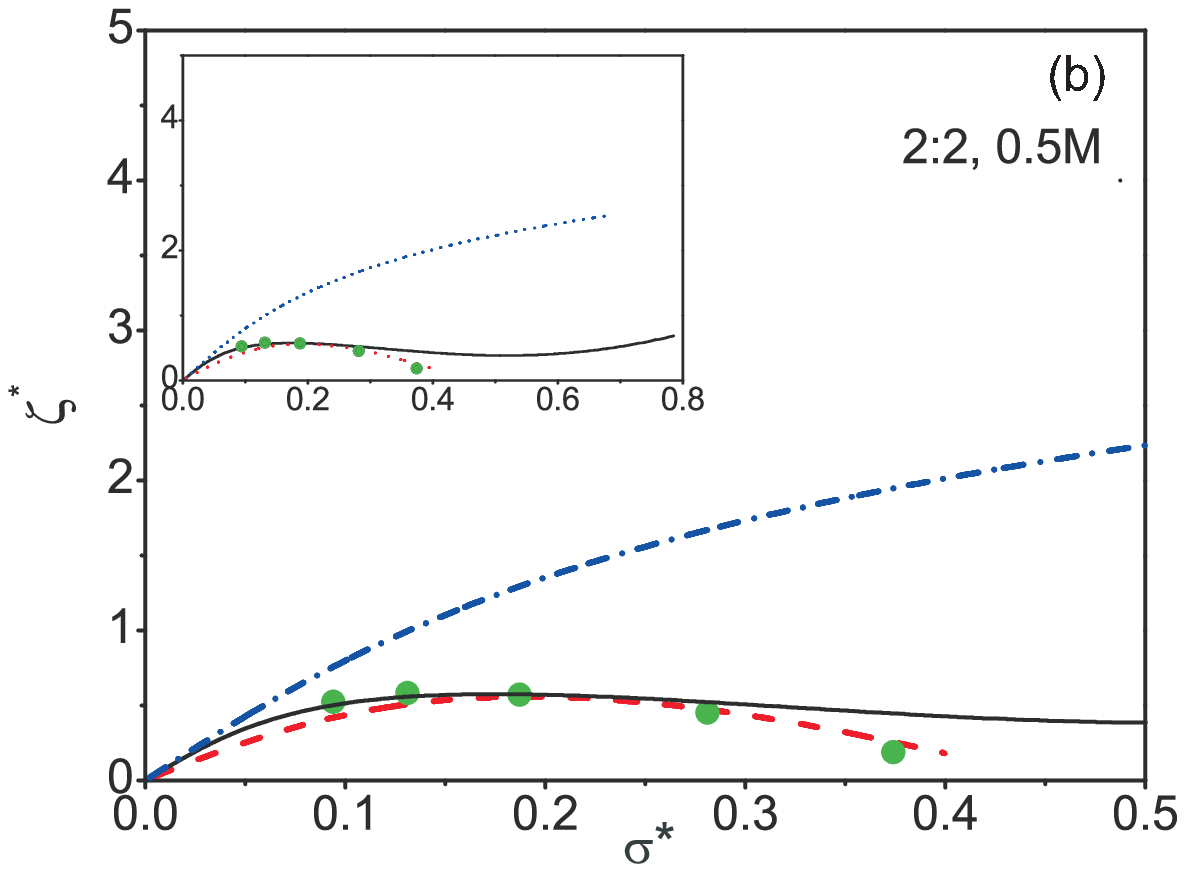}
}\vspace{1ex}
\centerline{
\includegraphics[width=0.45\textwidth]{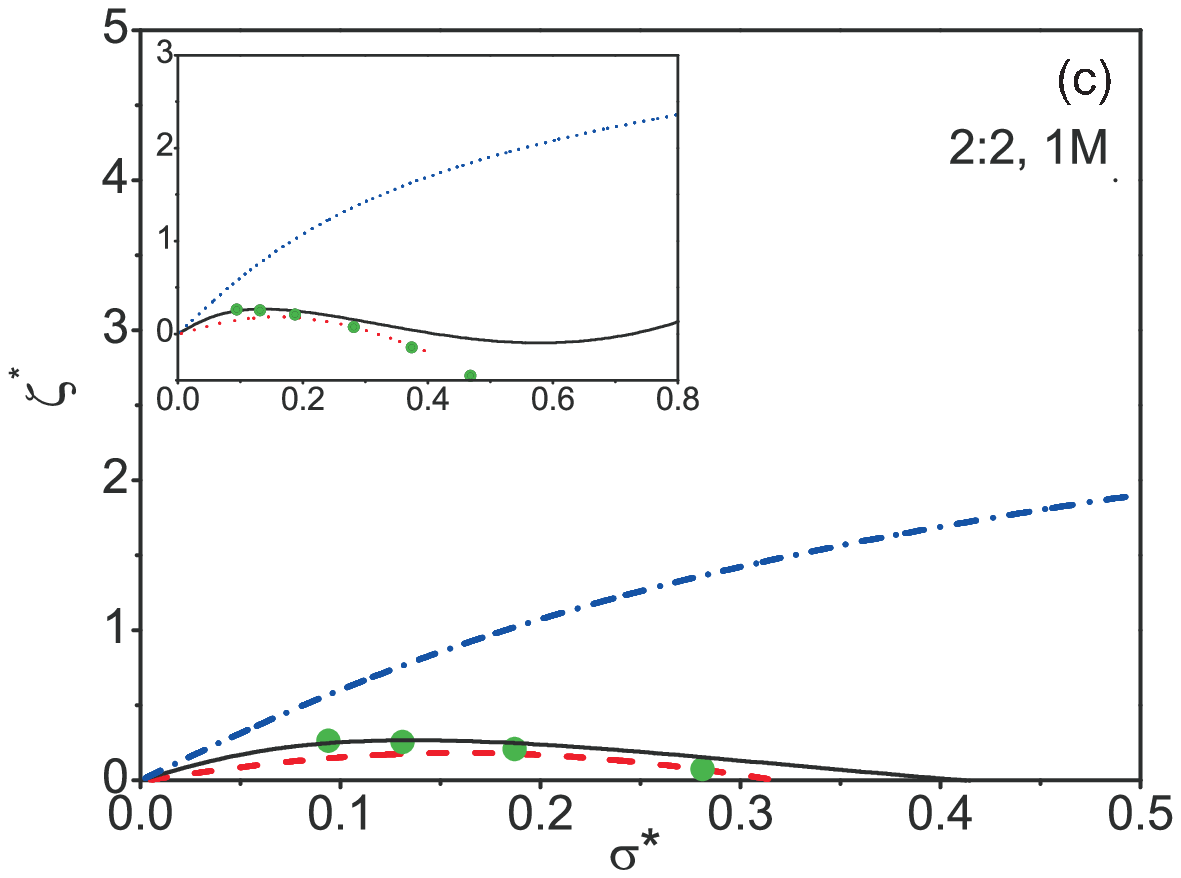}
}
\caption{(Color online) Reduced zeta potential $\zeta ^{*}$ [$=\psi ^{*}(R/a+1/2)$] versus
reduced surface charge density $\sigma ^{*}$ for a 2:2 electrolyte
at solution concentration (in mol/dm$^{3}$ units) 0.05~M (panel a),
0.5~M (panel b), and 1~M (panel c) in a RPM cylindrical double layer.
The symbols represent MC data, while the solid
line represents the MPB results, the dashed line represents the DFT results,
and the dash-dotted line shows the PB results. The polyion radius is  $8\times 10^{-10}$~m,
and the ionic diameter $a = 4 \times 10^{-10}$~m.
MC data from reference \cite{Goel}.}
\label{f2}
\end{figure}

\begin{figure}[!t]
\centerline{
\includegraphics[width=0.45\textwidth]{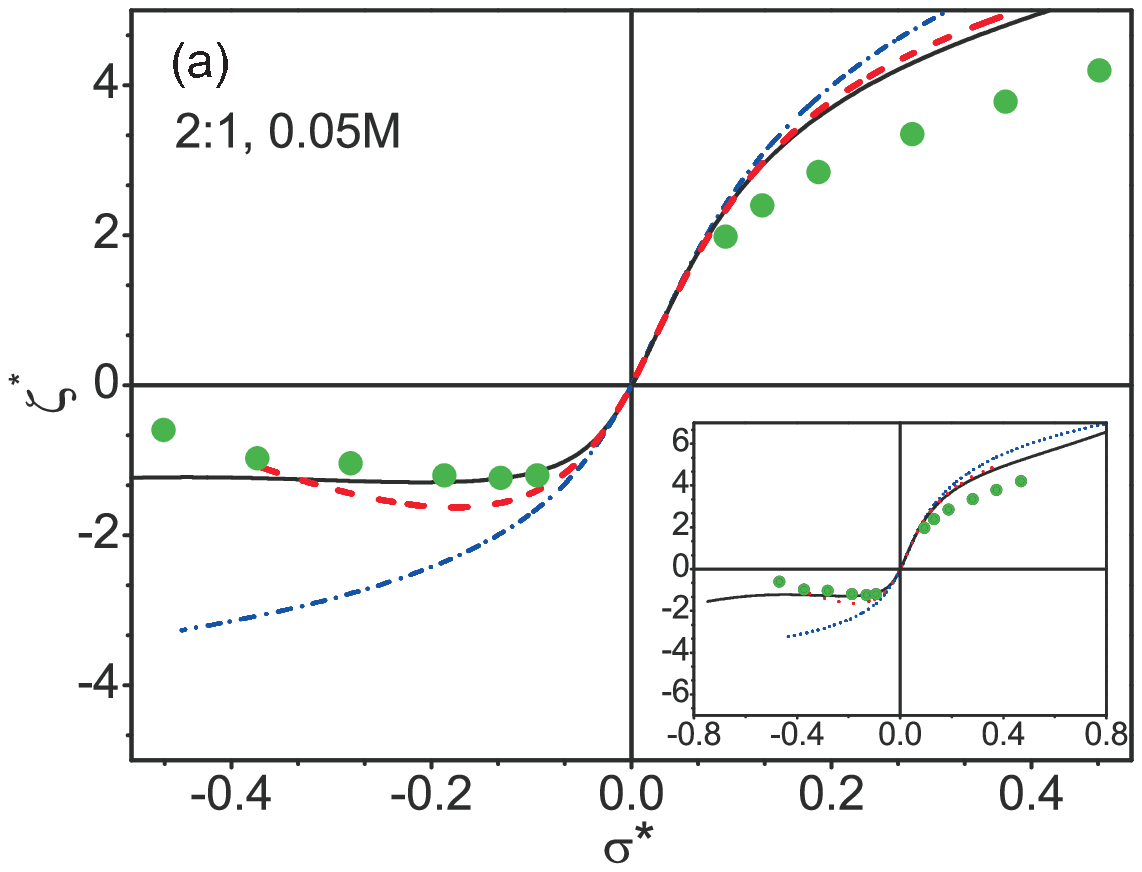}
\hspace{5mm}
\includegraphics[width=0.45\textwidth]{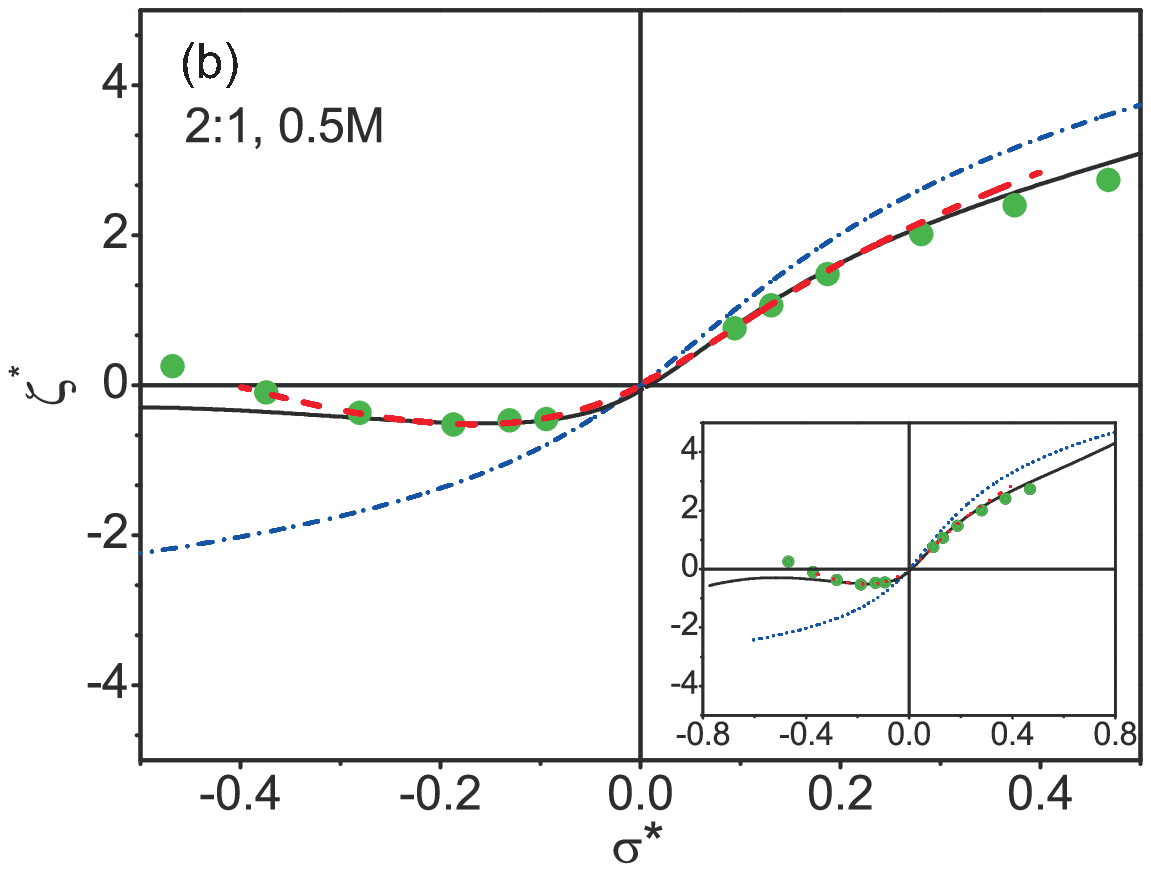}
}\vspace{1ex}
\centerline{
\includegraphics[width=0.45\textwidth]{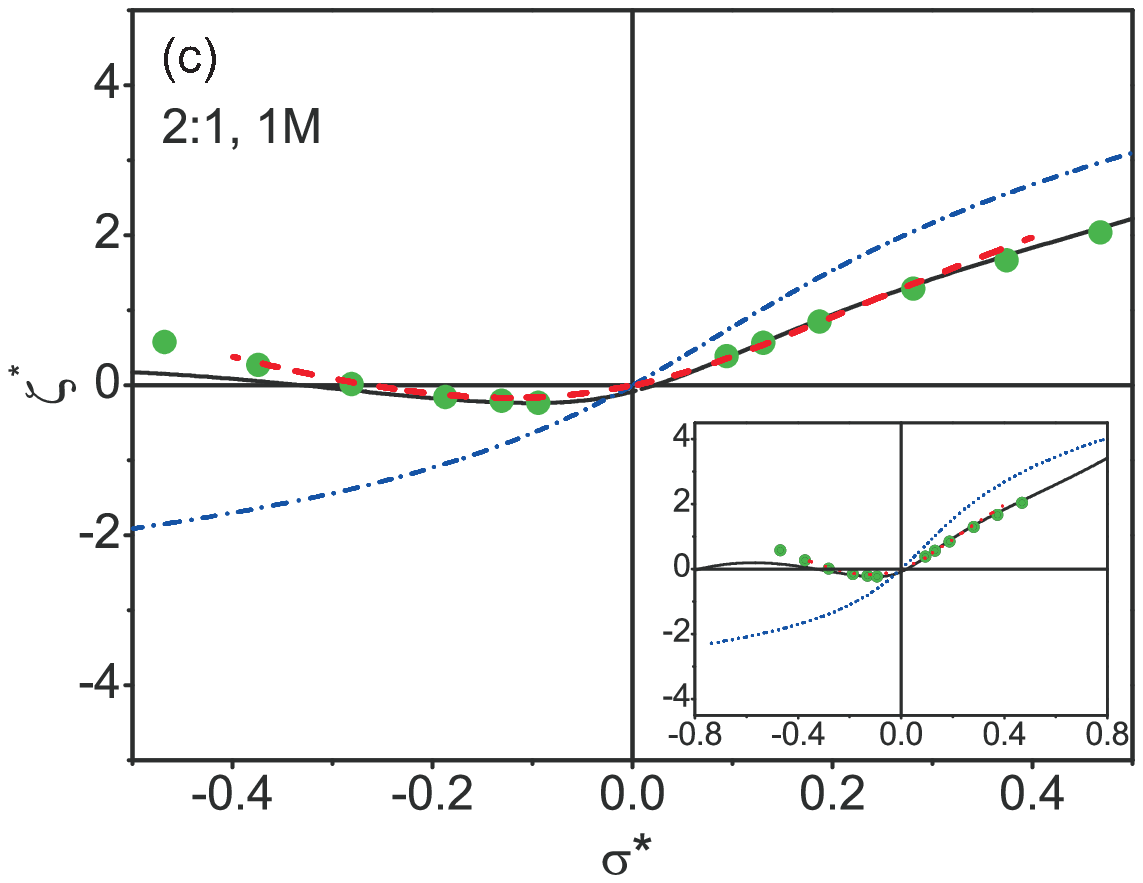}
}
\caption{(Color online) Reduced zeta potential $\zeta ^{*}$ [$=\psi ^{*}(R/a+1/2)$] versus
reduced surface charge density $\sigma ^{*}$ for a 2:1/1:2 electrolyte
at solution concentration (in mol/dm$^{3}$ units) 0.05~M (panel a),
0.5~M (panel b), and 1~M (panel c) in a RPM cylindrical double layer.
The symbols represent MC data, while the solid
line represents the MPB results, the dashed line represents the DFT results,
and the dash-dotted line shows the PB results. The polyion radius is  $8\times 10^{-10}$~m,
and the ionic diameter $a = 4 \times 10^{-10}$~m.
MC data from reference \cite {Goel}.}
\label{f3}
\end{figure}

\begin{figure}[!t]
\centerline{
\includegraphics[width=0.8\textwidth]{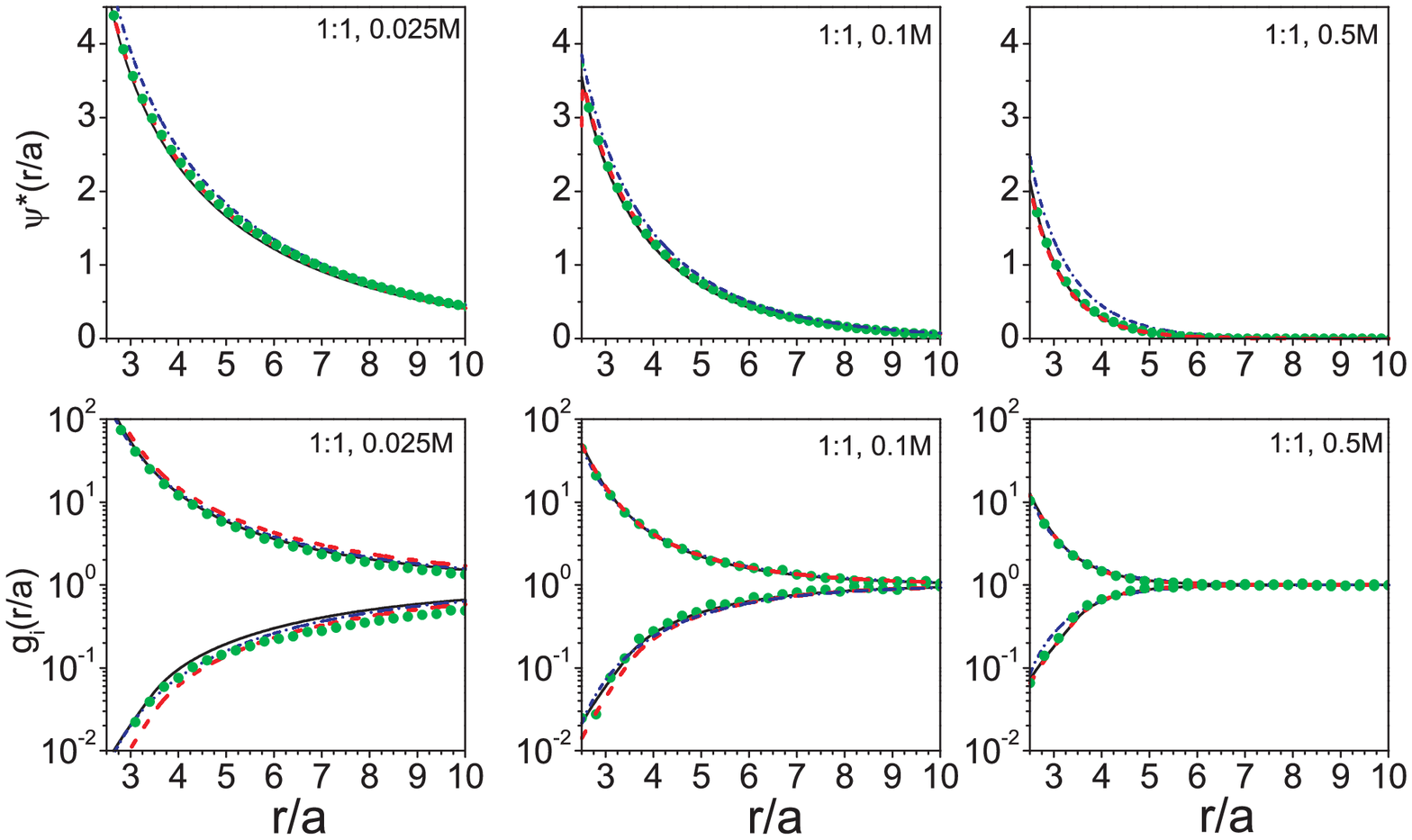}         
}
\caption{(Color online) Polyion-ion singlet distributions (lower panel) and the
reduced mean electrostatic potentials (upper panel) for a 1:1
electrolyte surrounding a polyion with an axial charge parameter $\xi = 4.2$
($\sigma = 0.187$~C/m$^{2}$) and radius $R = 8 \times 10^{-10}$~m.
The electrolyte concentrations are, from left to right, 0.025~M, 0.1~M, and 0.5~M, respectively.
The rest of symbols and notation as in figure~\ref{f1}.
The ionic diameter $a = 4 \times 10^{-10}$~m.
MC data from reference \cite{Goel}.}
\label{f4}
\end{figure}
\begin{figure}[!b]
\centerline{
\includegraphics[width=0.8\textwidth]{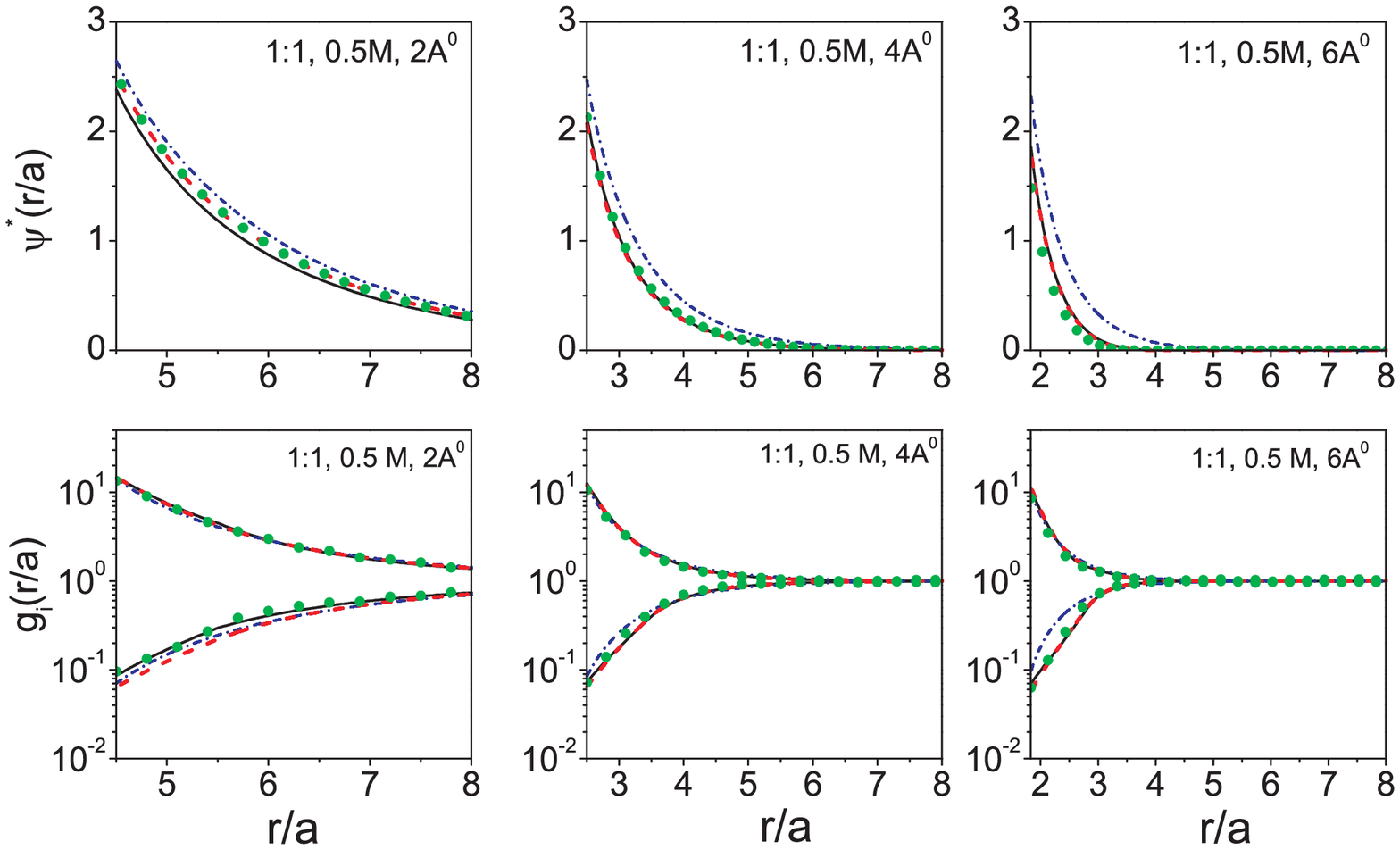}
}
\caption{(Color online) Polyion-ion singlet distributions (lower panel) and the
reduced mean electrostatic potentials (upper panel) for a 0.5~M 1:1
electrolyte surrounding a polyion with an axial charge parameter $\xi = 4.2$
($\sigma = 0.187$~C/m$^{2}$) and radius $R = 8 \times 10^{-10}$~m. The ionic diameters are,
from left to right, 2, 4, and  $6 \times 10^{-10}$~m (in the labels {\AA} $=  10^{-10}$~m), respectively.
The rest of symbols and notation as in figure~\ref{f1}.
MC data from reference \cite{Goel}.}
\label{f5}
\end{figure}

\begin{figure}[!t]
\centerline{
\includegraphics[width=0.8\textwidth]{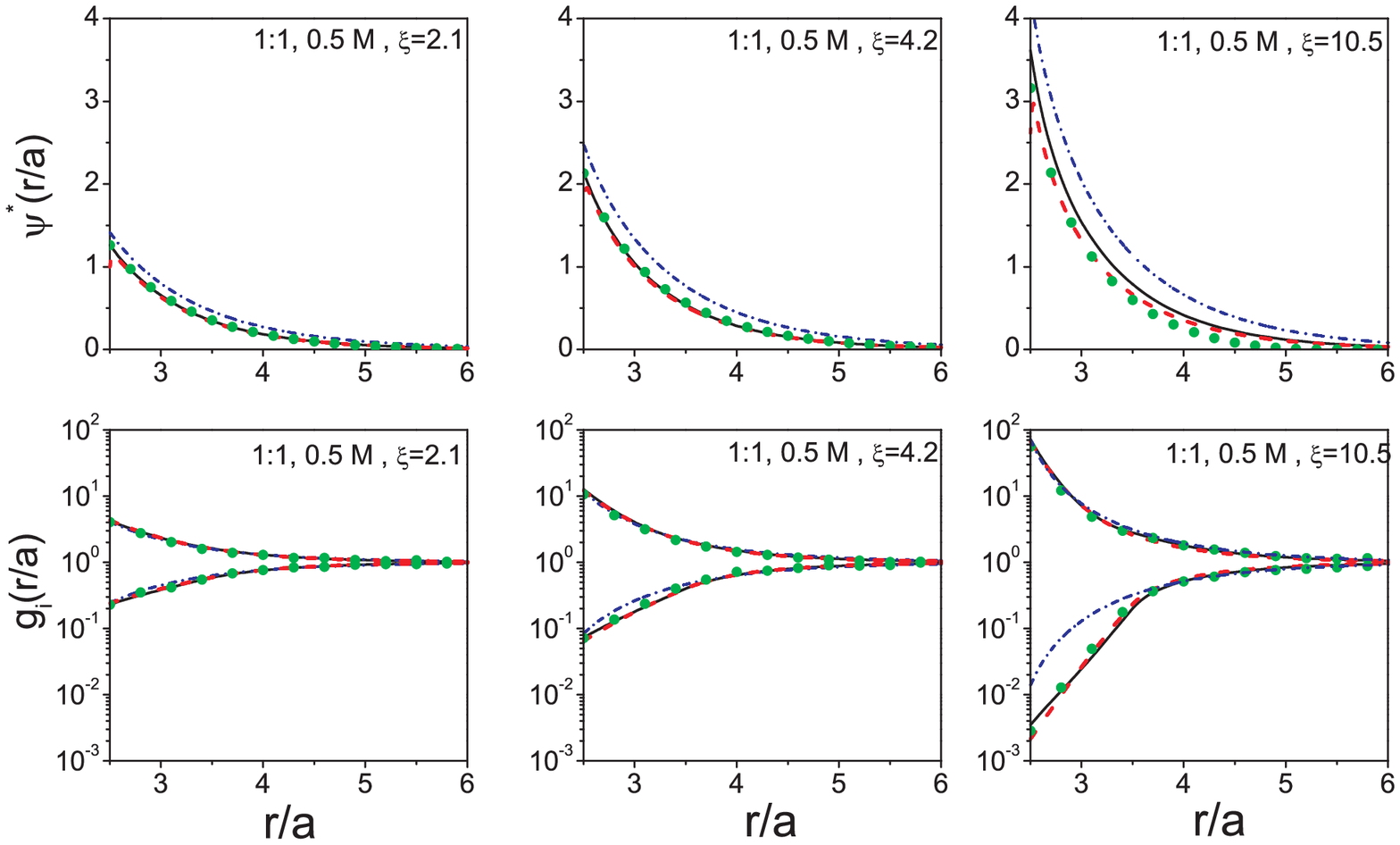}
}
\caption{(Color online) Polyion-ion singlet distributions (lower panel) and the
reduced mean electrostatic potentials (upper panel) for a 0.5~M 1:1
electrolyte surrounding a polyion with a radius $R = 8 \times 10^{-10}$~m. The
polyion axial charge parameter $\xi $ values are,
from left to right, 2.1 ($\sigma = 0.0935$~C/m$^{2}$), 4.2 ($\sigma = 0.187$~C/m$^{2}$),
and 10.5 ($\sigma = 0.468$~C/m$^{2}$), respectively. The ionic diameter is  $4\times 10^{-10}$~m
The rest of symbols and notation as in figure~\ref{f1}.
MC data from reference \cite{Goel}.}
\label{f6}
\end{figure}

\begin{figure}[!b]
\centerline{
\includegraphics[width=0.75\textwidth]{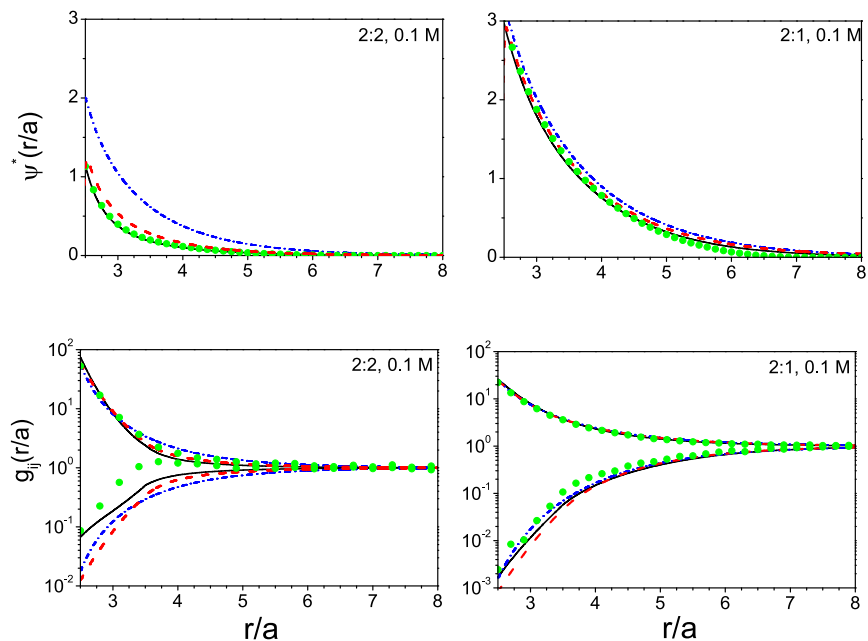}
}
\caption{(Color online) Polyion-ion singlet distributions (lower panel) and the
reduced mean electrostatic potentials (upper panel) for  2:2 and  2:1
electrolytes surrounding a polyion with an axial charge parameter $\xi = 4.2$
($\sigma = 0.187$~C/m$^{2}$) and radius $R = 8 \times 10^{-10}$~m. The electrolyte concentration
is 0.1 M in each case. The rest of symbols and notation as in figure~\ref{f1}.
The ionic diameter $a = 4 \times 10^{-10}$~m.}
\label{f7}
\end{figure}

\begin{figure}[!t]
\centerline{
\includegraphics[width=0.8\textwidth]{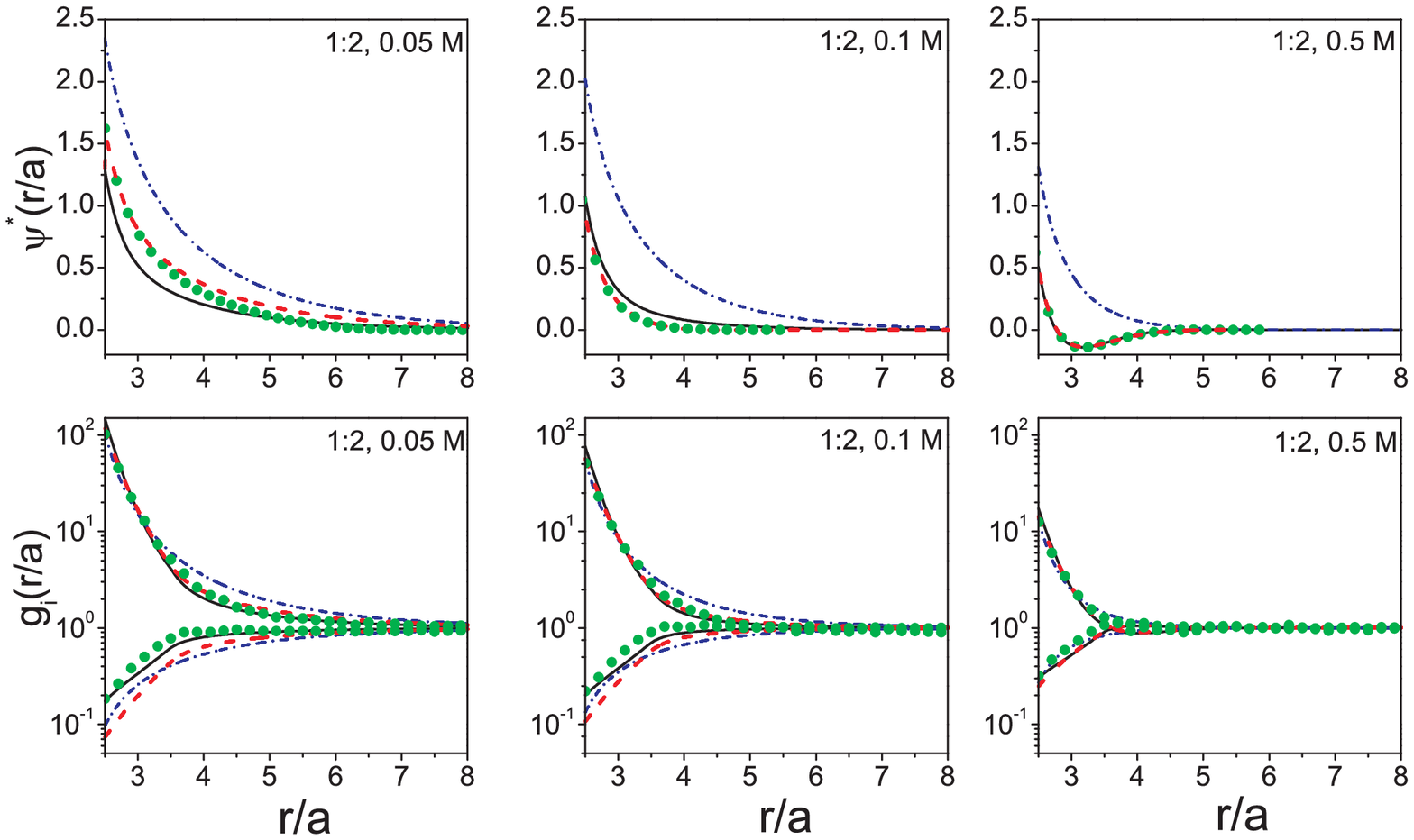}
}
\caption{(Color online) Polyion-ion singlet distributions (lower panel) and the
reduced mean electrostatic potentials (upper panel) for a 1:2
electrolyte surrounding a polyion with an axial charge parameter $\xi = 4.2$
($\sigma = 0.187$~C/m$^{2}$) and radius $R = 8 \times 10^{-10}$~m.
The electrolyte concentrations are, from left to right, 0.05~M, 0.1~M, and 0.5~M,
respectively. The rest of symbols and notation as in figure~\ref{f1}.
The ionic diameter $a = 4 \times 10^{-10}$~m.
MC data from reference \cite{Goel}.}
\label{f8}
\end{figure}

\begin{figure}[!b]
\centerline{
\includegraphics[width=0.8\textwidth]{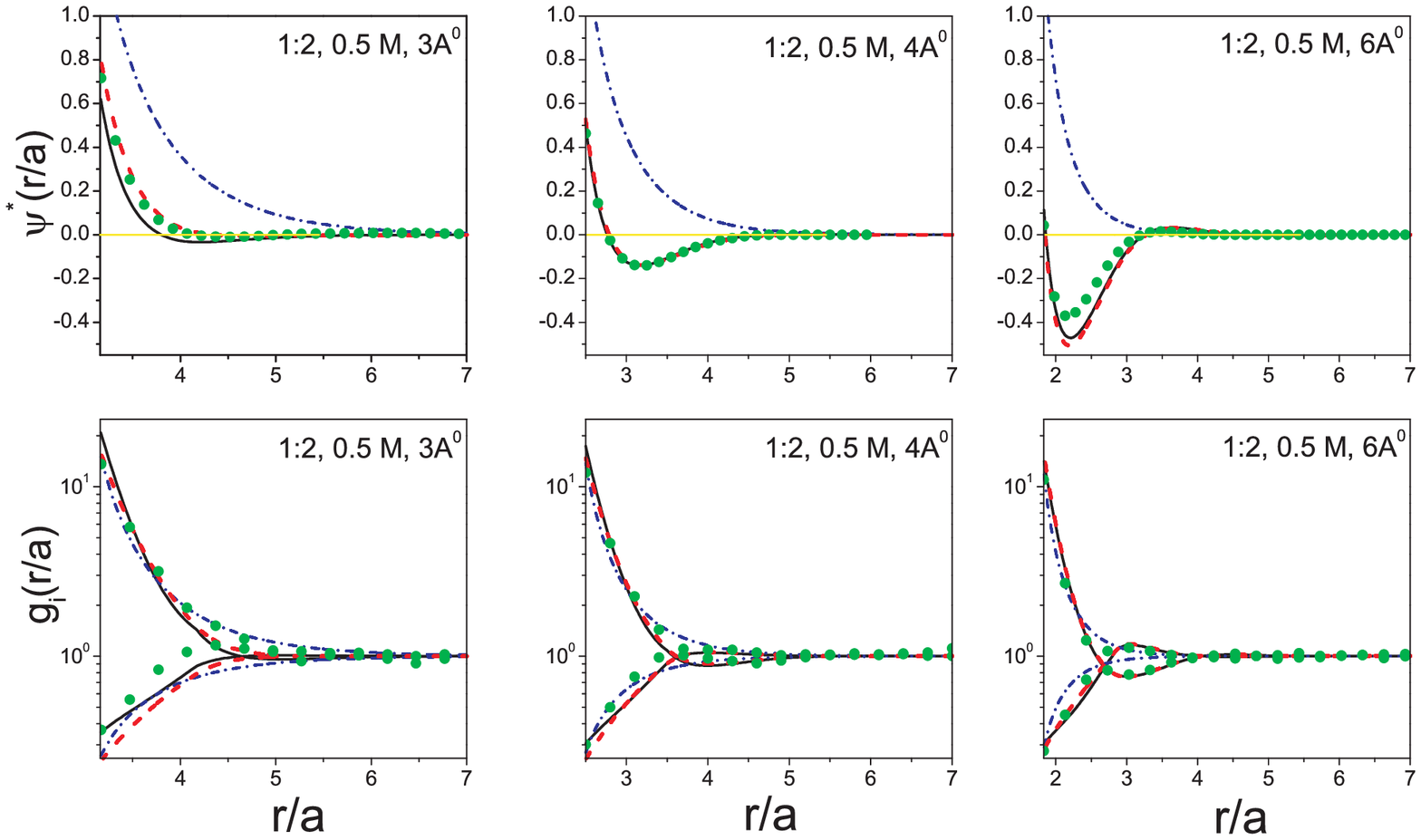}
}
\caption{(Color online) Polyion-ion singlet distributions (lower panel) and the
reduced mean electrostatic potentials (upper panels) for a 0.5~M 1:2
electrolyte surrounding a polyion with an axial charge parameter $\xi = 4.2$
($\sigma = 0.187$~C/m$^{2}$) and radius $R = 8 \times 10^{-10}$~m. The ionic diameters are,
from left to right, 3, 4, and  $6\times 10^{-10}$~m (in the labels {\AA} $=  10^{-10}$~m), respectively.
The rest of symbols and notation as in figure~\ref{f1}.
MC data from reference \cite{Goel}.}
\label{f9}
\end{figure}

\begin{figure}[!t]
\centerline{
\includegraphics[width=0.8\textwidth]{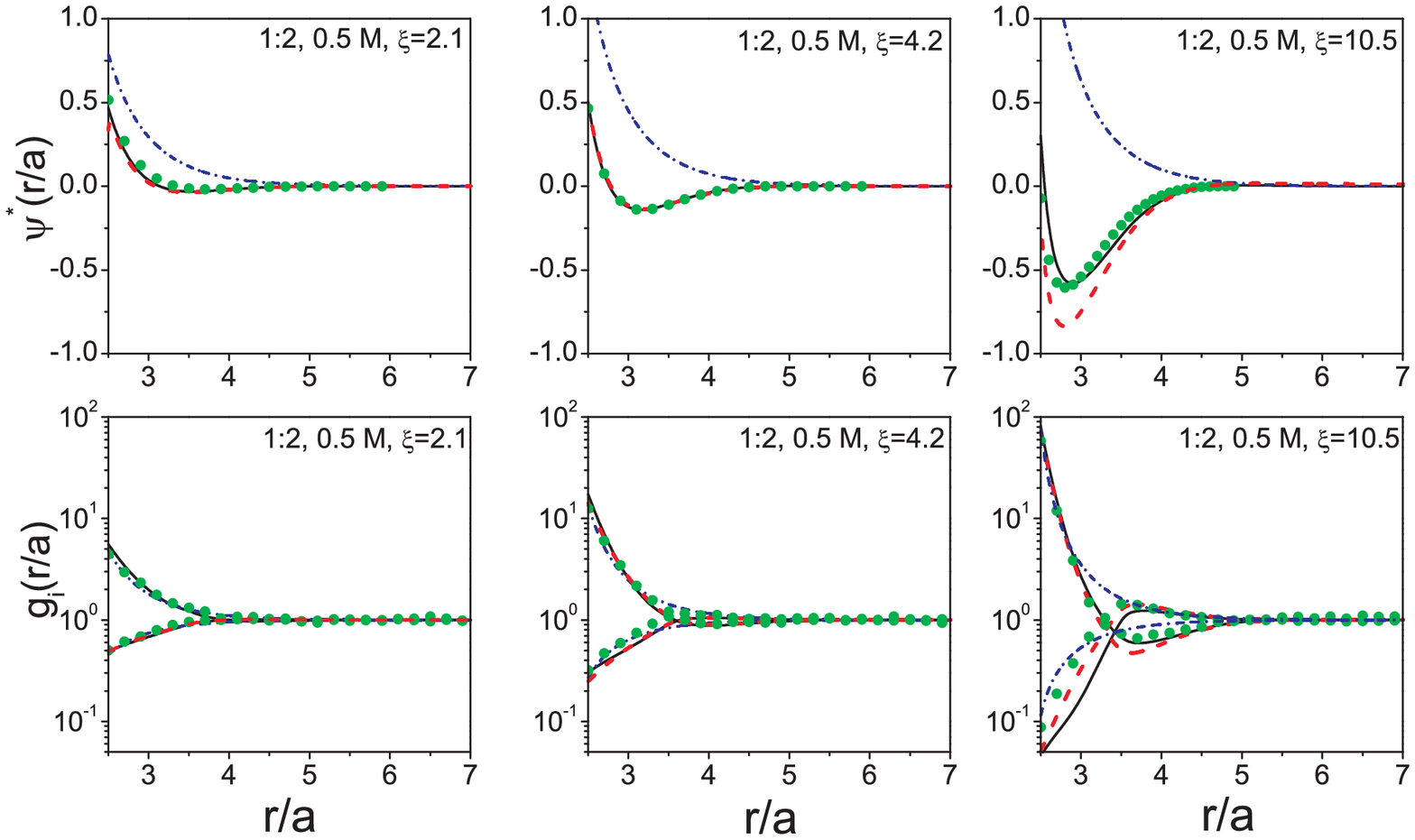}
}
\caption{(Color online) Polyion-ion singlet distributions (lower panel) and the
reduced mean electrostatic potentials (upper panel) for a 0.5~M 1:2
electrolyte surrounding a polyion with a radius $R = 8 \times 10^{-10}$~m. The
polyion axial charge parameter $\xi $ values are,
from left to right, 2.1 ($\sigma = 0.0935$~C/m$^{2}$), 4.2 ($\sigma = 0.187$~C/m$^{2}$),
and 10.5 ($\sigma = 0.468$~C/m$^{2}$), respectively. The ionic diameter is  $4 \times 10^{-10}$~m
The rest of symbols and notation as in figure~\ref{f1}.
MC data from reference \cite{Goel}.}
\label{f10}
\end{figure}

The DFT and MPB equations were solved numerically following
the procedure given in references \cite{ttdsw1,pg3} for DFT and \cite{bk,outh2,obl2}
for MPB. During the course of calculations we generally considered
the following ranges of variation for some of the physical parameters: the electrolyte
concentration $c$ from 0.05~mol/dm$^{3}$ to 2~mol/dm$^{3}$, the ionic diameter $a$
from  $2\times 10^{-10}$~m to  $6\times 10^{-10}$~m, and the axial charge factor $\xi$ from
2.1 ($\sigma = 0.0935$~C/m$^{2}$) to 10.5 ($\sigma = 0.468$~C/m$^{2}$). These parameters were
chosen to be in line with the existing MC data \cite{Goel}. It ought to be mentioned
though that the same ranges of variation for the same parameters were not strictly maintained
for all the electrolyte systems treated.  For comparison purposes we
have also obtained numerical solution of the classical PB theory at the same set of
physical states. In describing the results we will use universal reduced parameters
for convenience, the relevant ones in the present case being $\sigma^{\ast}(=\sigma d^{2}/e)$ for the reduced
surface charge density, and $\psi^{\ast}(r)$ [$= \beta e\psi(r)$] for the reduced mean electrostatic
potential.

\subsection{Zeta potential}

We begin this discussion by considering the zeta potential profile as a function
of polyion surface charge density $\sigma$, salt concentration $c$, and ionic valency $Z_{i}$.
The zeta potential is defined in the polyelectrolyte literature as the mean
electrostatic potential at the closest approach between
the small ion and the charged polyion, that is, $\zeta = \psi(R+a/2)$.
Figures~\ref{f1}--\ref{f3} depict the reduced zeta potential profiles $\zeta ^{*} = \psi ^{*}(R/a+1/2)$ with respect
to the reduced surface charge density $\sigma ^{*}$. In these calculations, the ionic diameter is
held fixed at $a = 4 \times 10^{-10}$~m. The results
for a 1:1 electrolyte system at three different electrolyte concentrations $c=0.1$~mol/dm$^{3}$,
1~mol/dm$^{3}$, and 2~mol/dm$^{3}$ are shown in figures~\ref{f1}~(a)--(c). In all cases the MC $\zeta ^{*}$ is monotonic and increases
continuously with $\sigma ^{*}$, although the rate of increment decreases at the higher salt concentration.
The PB results overestimate the  MC $\zeta ^{*}$ for the same $\sigma ^{*}$ and this deviation increases with concentration.
This is of course a well known feature of the mean field result in that the theory is more useful for monovalent
systems at low concentrations where the effect of the neglected inter-ionic correlations
is relatively less. One of the more noteworthy features of figure~\ref{f1} is the quantitative agreement of the DFT and
MPB predictions with MC results at all concentrations and over the whole range of $\sigma ^{*}$ studied.
A similar level of agreement between the two approaches was seen for the PDL
\cite{bo3} and the SDL \cite{bo4}.

As we move on to 2:2 systems in figure~\ref{f2}, the striking feature is that the MC, DFT, and MPB $\zeta ^{*}$
are no longer monotonic. Indeed, the MC $\zeta ^{*}$ shows a maximum before starting to decrease at
high $\sigma ^{*}$. Both DFT and MPB $\zeta ^{*}$ follow the MC trends closely. At still higher
surface charges, viz., $\sigma ^{*} \gtrsim  0.5$, the MPB shows a shallow valley before increasing
again as can be seen in the insets. This trend is probably an artefact of the theory. The PB results are monotonic
and are thus not even qualitative with the simulations.

Figure~\ref{f3} shows the $\zeta ^{*}$ for an asymmetric 2:1/1:2 valency electrolyte. In the
region $\sigma ^{*} >$ 0 we have univalent  counterions, while in the region $\sigma ^{*} <$ 0
the counterions are divalent. It is interesting to observe that the characteristics of
the zeta potential for the 2:1 and 1:2 systems resemble that for 1:1 in figure~\ref{f1}, and
2:2 in figure~\ref{f2}, respectively. This is clearly indicative of the well known result
that it is the electrode-counterion interaction that governs double layer properties.
Although not clearly visible at the scale of the figures, the MC, DFT, and MPB reveal
a non-zero potential of zero charge (pzc), which is a consequence of asymmetry in the system.
Besides, both DFT and MPB show good overall agreements with the simulations.
The PB theory, however, does not reveal any non-zero pzc and is again not qualitative with the MC results for 1:2
systems. We remark further that the maxima or minima in $\zeta ^{*}$ have been predicted previously
theoretically and through simulations for the PDL
\cite{bo3,bfhs1,ct1,bhpf1} and the SDL \cite{ywg1,bo4,goel2,dlsg1,dl1},
and have also been observed earlier for the CDL
\cite{bo1,bo2,lc1,glh1}.

\subsection{Double layer structure}

\subsubsection{1:1 electrolytes}

We now focus on the static structural
properties of the CDL given in figures~\ref{f4}--\ref{f10}.  In the three figures~\ref{f4}, \ref{f5}, and \ref{f6}
we will compare the
theoretical results with the available MC data \cite{Goel} for ionic
density distributions as well as the mean electrostatic potentials for 1:1 salts at different parametric conditions.
Figure~\ref{f4} depicts variation in concentration of the salt while keeping all other parameters fixed,
and in particular, $a = 4 \times 10^{-10}$~m and $\xi = 4.2$ ($\sigma = 0.187$~C/m$^{2}$).
As might be expected, there is an excessive accumulation of counterions near the polyion surface, which
continuously decays down to the bulk. All the $\psi ^{*}(r)$ and $g_{i}(r)$ are monotonically decreasing
with the DFT and MPB results being virtually indistinguishable from the MC results.
With an increase in electrolyte concentration, this decrease becomes more rapid and
the double layer becomes progressively less diffuse. An increase of concentration
also leads to an increase in coion contact value and to a decrease in counterion contact value. This is due to
a more complete screening at these conditions.
As has been seen previously in case of the PDL \cite{bo3}, at such relatively low
concentration 1:1 systems, the PB results are also in good agreement
with the simulations.

The effect of ionic size on the double layer
structure is illustrated in figure~\ref{f5}, where we present the results
at ionic diameters $a= 2$, $4$, and $6 \times 10^{-10}$~m, respectively.
These calculations are at $c= 0.5$~mol/dm$^{3}$, and $\xi = 4.2$
($\sigma = 0.187$~C/m$^{2}$). A glance at the figures across the panels
from left to right reveals that the double layer becomes more compact
as the ionic diameter increases. The consequent increase of the ion exclusion volume,
or equivalently the solute volume fraction, results in increased packing effects, which
reduces the range of the singlet distribution functions. This feature is corroborated
by the corresponding potential profiles. In figure~\ref{f6} the effect of charge
correlations on structure can be seen in the profiles now at different axial
charge parameters, viz., $\xi= 2.1$, $4.2$, and $10.5$, corresponding to $\sigma = 0.0935$, $0.187$,
and $0.468$~C/m$^{2}$. Here, the ionic diameter and concentration are held fixed
at  $4\times 10^{-10}$~m and 0.5~mol/dm$^{3}$, respectively. A high surface charge
density leads to stronger electrostatic correlations, which in turn leads again to compact
double layers. In effect, both size and charge correlations
affect the structure in the same directions. In the last two figures the
DFT and MPB reproduce the simulation results to a very good degree, while
the PB shows the maximum deviation at $a = 6 \times 10^{-10}$~m (figure~\ref{f5})
and $\xi=  10.5$ ($\sigma = 0.468$~C/m$^{2}$), figure~\ref{f6}.

\subsubsection{2:2 and 2:1/1:2 electrolytes}

Turning now to higher and/or multivalent electrolyte systems, in figure~\ref{f7} we present
the density and the potential profiles for 2:2 and 2:1 electrolytes
at 0.1~mol/dm$^{3}$ concentration, $a = 4 \times 10^{-10}$~m, and
$\xi = 4.2$ ($\sigma = 0.187$~C/m$^{2}$). We remark here that we have performed the
MC simulations for these cases in the course of this work as this data in
reference \cite{Goel} has been found to be somewhat doubtful. The 2:2 electrolyte with the divalent
counterion shows somewhat larger structure relative to the 2:1 electrolyte due to an increased
polyion-counterion attraction. This is also visible in the potential profiles
with the double layer being more diffuse in the latter case. The DFT and MPB results
again show a considerable consistency for both 2:2 and 2:1 salts, although
in the former instance the MPB coion $g(r)$ is marginally closer to the MC data near contact.
In the same case, the PB also shows a greater deviation.

   The role of an increased electrostatic attraction between the polyion and
multivalent counterions in characterizing the
polyion--ion distributions and mean electrostatic potential
profiles is further evident in figures~\ref{f8}--\ref{f10}, which
show the structure of a double layer for 1:2 electrolytes
under different physical conditions. Figure~\ref{f8} shows the effect of
an increasing concentration at fixed $a = 4 \times 10^{-10}$~m and
$\xi = 4.2$ ($\sigma = 0.187$~C/m$^{2}$). While at the
two lower concentrations the profiles are all monotonic, at the
highest concentration ($c =  0.5$~mol/dm$^{3}$) oscillations are visible
in the $g_{i}$'s  and $\psi ^{*}$. In particular, the minimum
in $\psi ^{*}$ next to the polyion indicates an overscreening (charge reversal)
of the polyion by the counterions. This has been observed in
earlier double layer studies involving,
besides the DFT and MPB, other theoretical and numerical methods
\cite{yfsl1}. The DFT and MPB theories follow the MC data very closely,
whereas the classical theory shows a substantial deviation especially
at $c = 0.5$~mol/dm$^{3}$ where its monotonic behaviour is at odds with that
of the simulations.

Overscreening can also be a function of ionic size as well the
surface charge density of the polyion as can be gleaned from figures~\ref{f9} and \ref{f10}.
A rather large charge inversion is observed in figure~\ref{f9}
when the depth of the potential minimum increases with an increase in ionic size.
The comparative behaviour of the DFT and the MPB theories relative to the
MC simulations seen earlier extends to these situations too, while
the PB theory continues to show large and sometimes qualitative discrepancies.
Charge correlation shows its prominence in figure~\ref{f10} with overscreening leading to
large charge inversion at the highest surface charge density treated
[$\xi = 10.5$ ($\sigma = 0.486$~C/m$^{2}$)]. However, in this situation the DFT
coion distribution shows some deviation from the MC and MPB results, similar
to that for the 2:2 case in figure~\ref{f7}, possibly due
to the approximation involved in the present formulation of the theory.
The DFT charge inversion is also overestimated at $\xi = 10.5$.
In the classical formalism, steric effects or charge correlations are never taken into
account and as such the PB theory yields monotonically decreasing density and
potential profiles at all concentrations, ionic sizes, and surface charge densities.

\section{Conclusions}

We have done a comparative study of the density functional and
the modified Poisson-Boltzmann theories as applied to the RPM
cylindrical double layer. A critical evaluation of these analytical approaches
in their ability to reproduce the Monte Carlo results for the zeta potentials,
the singlet density distributions, and the mean electrostatic potential profiles
has been the theme of this work. An important global aspect of the results is the
consistency of both theories in their predictions of these structural
features for the range of the concentrations of the electrolyte, ionic diameters,
and the polyion surface charge densities studied here; the theories also
follow the Monte Carlo simulations to a very good degree overall.

The classical mean field results are reasonable for 1:1 valencies
at relatively low concentrations, but are generally poor and  even fail to be
qualitative with the simulations and the formal theoretical results at
higher concentrations and/or in the presence of higher valency
or asymmetric valency electrolytes. These results are not unsurprising and
point to the importance of including ionic correlations and exclusion volume effects.
The appearance of a maximum (or minimum) in the zeta potentials for
electrolytes with multivalent counterions is an interesting phenomenon
and is suggestive of a non-monotonic nature of the differential capacitance.
In such situations, the density and potential profiles also indicate
polyion overcharging \cite{bo4,goel2}.

The present study of the double layer in cylindrical geometry in
conjunction with the earlier studies in planar \cite{bo3}, and spherical
\cite{bo4} geometries reveal remarkable overall consistency of the DFT and MPB theories
both among themselves and with the corresponding simulation data. and hence the
usefulness of these theories in describing interfacial double layers. This is
encouragement for a future development and application of these theories to
more complex polyelectrolyte and colloidal systems.

\section*{Acknowledgements}

V. D. acknowledges an institutional grant through FIPI, University of
Puerto Rico during the initial stages of this project. C.N.P would like
to thank Drs. T. Goel and S.K. Ghosh for useful discussions.


\ukrainianpart

\title
{Структура циліндричного подійного шару. Порівняння теорій
функціоналу густини і модифікованого рівняння Пуассона-Больцмана з
моделюванням Монте-Карло}

\author{В. Дорвільєн\refaddr{label1}, К.Н. Патра\refaddr{label2},
Л.Б. Буйан\refaddr{label1}, К.В. Оутвайт\refaddr{label3}}

\addresses{
\addr{label1}Лабораторія теоретичної фізики, Університет Пуерто
Ріко, Пуерто Ріко, США \addr{label2} Відділ теоретичної хімії, Центр
атомних досліджень Бгабга, Мумбай, Індія \addr{label3} Факультет
прикладної математики, Університет м. Шеффілда,  Шеффілд,
Великобританія }

\makeukrtitle

\begin{abstract}
\tolerance=3000%
Вивчено структуру циліндричних подвійних шарів, використовуючи
модифіковану теорію Пуассона-Больцмана і теорію функціоналу густини.
У цій моделі подвійного  шару електродом є нескінченно довгий,
непроникний та однорідно заряджений циліндричний полііон. Він
оточений системою іонів однакового розміру у діелектричному
середовищі. Виконано докладне порівняння результатів теорії для
дзета потенціалів, для середнього електростатичного потенціалу та
для  розподілів густини іон-електрод з даними комп'ютерного
моделювання  Монте Карло. Теоретичні результати узгоджуються з
новими та попередніми даними моделювання Монте Карло при зміні
різних параметрів, таких як іонні діаметри, концентрація електрода,
густина заряду електрода.

\keywords подвійний електричний шар, обмежена примітивна модель,
профілі густини
\end{abstract}


\begin{thebibliography}{99}

\bibitem{mandel1} Mandel~M.,  Polyelectrolytes, Reidel, Dordrecht, 1988.

\bibitem{hara1} Hara M., Polyelectrolytes: Science and Technology, Dekker, New York, 1993.

\bibitem{ha1} Harrison S.C., Aggarwal A.K.,  Annu. Rev. Biochem., 1990, \textbf{59}, 933; \doi{10.1146/annurev.bi.59.070190.004441}.


\bibitem{mhsfh1} Mishra V.K., Hecht J.L., Sharp K.A., Friedman R.A., Honig B.,  J. Mol. Biol., 1994, \textbf{238,} 264; \\ \doi{10.1006/jmbi.1994.1286}.

\bibitem{js1} Jary D., Sikorav J.-L.,  Biochemistry, 1999, \textbf{28}, 3223; \doi{10.1021/bi982770h}.


\bibitem{hb}  Henderson D.,  Boda D.,  Phys. Chem. Chem. Phys.,  2009, \textbf{11}, 3822;
\doi{10.1039/b815946g}.

\bibitem{cher} Chersty A.G., Phys. Chem. Chem. Phys., 2011, \textbf{13}, 9942;
\doi{10.1039/c0cp02796k}.

\bibitem{bov96} Bhuiyan L.B., Outhwaite C.W., van der Maarel J.R.C.,  Physica A, 1996, \textbf{231}, 295; \doi{10.1016/0378-4371(95)00443-2}.

\bibitem{zebobvm} Zakharova S.S., Egelhaaf S.U., Bhuiyan L.B., Outhwaite C.W.,
Bratko B., van der Maarel J.R.C.,  J.~Chem. Phys., 1999, \textbf{111}, 10706; \doi{10.1063/1.480425}.

\bibitem{isr1} Israelachvili J., Intermolecular and Surface Forces, Academic Press, New York, 1992.

\bibitem{lr1} Imaging of Surfaces and Interfaces, Lipkowski J., Ross P. (Eds.), Wiley-VCH, New York, 1999.

\bibitem{levin} Levin Y., Rep. Prog. Phys., 2002, \textbf{65}, 1577; \doi{10.1088/0034-4885/65/11/201}.

\bibitem{mann2} Manning G.S.,  Acc. Chem. Res., 1979, \textbf{12}, 443;
\doi{10.1021/ar50144a004}.

\bibitem{stigter1} Stigter D.,  J. Phys. Chem., 1978, \textbf{82}, 1603;
\doi{10.1021/j100503a006}.

\bibitem{sh1} Sharp K.A., Honig B.,  J. Phys. Chem., 1990, \textbf{94}, 7684; \doi{10.1021/j100382a068}.

\bibitem{hend1} Fundamentals of Inhomogeneous Fluids, Henderson D. (Ed.), Dekker, New York, 1992.

\bibitem{yfsl1} Yeomans L., Feller S.E., S\'{a}nchez E., Lozada-Cassou M.,  J. Chem. Phys., 1993, \textbf{98}, 1436;
\doi{10.1063/1.464308}.

\bibitem{outh} Outhwaite C.W.,  J. Chem. Soc. Faraday Trans. 2,  1986,  \textbf{82}, 789; \doi{10.1039/f29868200789}.

\bibitem{bo1} Bhuiyan L.B., Outhwaite C.W., In: Condensed Matter Theories, Vol.~8, Blum L., Malik F.B. (Eds.), New York, Plenum, 1993, 551--555.

\bibitem{bo2} Bhuiyan L.B., Outhwaite C.W.,  Philos. Mag. B, 1994, \textbf{69}, 1051; \doi{10.1080/01418639408240174}.

\bibitem{py1} Patra C.N., Yethiraj A.,  J. Phys. Chem. B, 1999,
\textbf{103}, 6080; \doi{10.1021/jp991062i}.

\bibitem{py2} Patra C.N., Yethiraj A.,  Biophys. J., 2000,
\textbf{78}, 699; \doi{10.1016/S0006-3495(00)76628-8}.

\bibitem{mar1} Mills P., Anderson C.F., Record M.T. (Jr.), J. Phys. Chem., 1985, \textbf{89}, 3984; \doi{10.1021/j100265a012}.

\bibitem{mar2} Mills P., Anderson C.F., Record M.T. (Jr.), J. Phys. Chem., 1986, \textbf{90}, 6541; \doi{10.1021/j100282a025}.

\bibitem{nar1} Ni H.H., Anderson C.F., Record M.T. (Jr.),  J. Phys. Chem. B, 1999, \textbf{103}, 3489; \doi{10.1021/jp984380a}.

\bibitem{ar1} Anderson C.F., Record M.T. (Jr.),  Annu. Rev. Phys. Chem., 1982, \textbf{33}, 191; \doi{10.1146/annurev.pc.33.100182.001203}.


\bibitem{dbbo1} Das T., Bratko D., Bhuiyan L.B., Outhwaite C.W.,  J. Phys. Chem., 1995, \textbf{99}, 410; \doi{10.1021/j100001a061}.


\bibitem{dbbo2} Das T., Bratko D., Bhuiyan L.B., Outhwaite C.W.,  J. Chem. Phys., 1997, \textbf{107}, 9197; \doi{10.1063/1.475211}.

\bibitem{bv1} Bratko D., Vlachy V.,  Chem. Phys. Lett., 1982, \textbf{90} 434; \doi{10.1016/0009-2614(82)80250-9}.

\bibitem{bv2} Bratko D., Vlachy V.,  Chem. Phys. Lett., 1985, \textbf{115}, 294; \doi{10.1016/0009-2614(85)80031-2}.

\bibitem{lbz} Le Bret M., Zimm B.H.,  Biopolymers, 1984, \textbf{23}, 271; \doi{10.1002/bip.360230208}.

\bibitem{vh1} Vlachy V., Haymet A.D.J.,  J. Chem. Phys., 1986, \textbf{84},
5874; \doi{10.1063/1.449898}.

\bibitem{pb1} Patra C.N., Bhuiyan L.B.,  Condens. Matter Phys., 2005, \textbf{8}, 425; \doi{10.5488/CMP.8.2.425}.

\bibitem{Goel} Goel T., Patra C.N., Ghosh S.K.,  Mukherjee T.,  J. Chem. Phys., 2008,
\textbf{129}, 154906; \doi{10.1063/1.2992525}.

\bibitem{bo3} Bhuiyan L.B., Outhwaite C.W., Phys. Chem. Chem. Phys., 2004,
\textbf{6}, 3467; \doi{10.1039/b316098j}.

\bibitem{bo4} Bhuiyan L.B., Outhwaite C.W.,  Condens. Matter Phys., 2005, \textbf{8},
287; \doi{10.5488/CMP.8.2.287}.

\bibitem{goel2} Goel T., Patra C.N.,  J. Chem. Phys., 2007, \textbf{127}, 034502;
\doi{10.1063/1.2750335}.

\bibitem{boh1} Bhuiyan L.B., Outhwaite C.W.,  Henderson D.,  J. Chem. Phys., 2005, \textbf{123}, 034704; \doi{10.1063/1.1992427}.

\bibitem{ttdsw1} Tang Z., Mier-y-Teran L., Davis H.T., Scriven L.E., White H.S.,
Mol. Phys., 1990, \textbf{71} 369; \\ \doi{10.1080/00268979010001851}.


\bibitem{tswd1} Mier-y-Teran L., Suh S.H., White H.S., Davis
H.T.,  J. Chem. Phys., 1990, \textbf{92}, 5087;
\doi{10.1063/1.458542}.

\bibitem{ttdsw2} Mier-y-Teran L., Tang Z., Davis H.T., Scriven L.E., White H.S.,  Mol. Phys., 1991, \textbf{72}, 817;
\\ \doi{10.1080/00268979100100581}.

\bibitem{lw06} Li Z., Wu J.,  Phys. Rev. Lett., 2006, \textbf{96}, 048302;
\doi{10.1103/PhysRevLett.96.048302}.

\bibitem{pg3} Patra C.N.,  Ghosh S.K.,  J. Chem. Phys., 1994, \textbf{100}, 5219; \doi{10.1063/1.467186}.

\bibitem{da1} Denton A.R.,  Ashcroft N.W.,  Phys. Rev. A, 1989, \textbf{39}, 426; \doi{10.1103/PhysRevA.39.426}.

\bibitem{lc1} Lozada-Cassou M.,  J. Phys. Chem., 1983, \textbf{87}, 3729;
\doi{10.1021/j100242a031}.

\bibitem{glh1} Gonzlaez-Tovar E., Lozada-Cassou M., Henderson D.,  J. Chem. Phys., 1985,
\textbf{83}, 361; \doi{10.1063/1.449779}.

\bibitem{bfhs1} Boda D., Fawcett W.R., Henderson D., Sokolowski S.,  J. Chem. Phys., 2002,
\textbf{116}, 7170; \doi{10.1063/1.1464826}.

\bibitem{tn1} Terao T., Nakayama N.,  Phys. Rev. E, 2001, \textbf{63}, 041401;
\doi{10.1103/PhysRevE.63.041401}.

\bibitem{bk} Bellman R., Kalaba R., Quasilinearization and Nonlinear Boundary Value Problems,  Elsevier, New York, 1965.

\bibitem{outh2} Outhwaite C.W., J. Chem. Soc. Faraday Trans. 2, 1987, \textbf{83}, 949;
\doi{10.1039/F29878300949}.

\bibitem{obl2} Outhwaite C.W., Bhuiyan L.B., Levine S.,  J. Chem. Soc. Faraday Trans. 2, 1980, \textbf{76}, 1388;
\doi{10.1039/F29807601388}.

\bibitem{ct1} Carnie S.L.  Torrie G.M.,  Adv. Chem. Phys., 1984, \textbf{56}, 141;
\doi{10.1002/9780470142806.ch2}.

\bibitem{bhpf1} Boda D., Henderson D., Plaschko P., Fawcett W.R.,  Mol. Simulat., 2004,
\textbf{30}, 137; \doi{10.1080/0892702031000152163}.

\bibitem{ywg1} Yu Y.-X., Wu J.  Gao G.-H.,  J. Chem. Phys., 2004, \textbf{120}, 7223;
\doi{10.1063/1.1676121}.

\bibitem{dlsg1} Degreve L., Lozada-Cassou M., Sanchez E., Gonzalez-Tovar E.,  J. Chem. Phys., 1993, \textbf{98}, 8905;
\doi{10.1063/1.464449}.

\bibitem{dl1} Degreve L., Lozada-Cassou M.,  Mol. Phys., 1995, \textbf{86} 759;
\doi{10.1080/00268979500102351}.

\end{thebibliography}
\end{document}